# Combination of binaural and harmonic masking release effects in the detection of a single component in complex tones

Martin Klein-Hennig, Mathias Dietz, and Volker Hohmann[a]

Medizinische Physik and Cluster of Excellence Hearing4all, Universität Oldenburg, 26111 Oldenburg, Germany.

a) Corresponding author, electronic mail: volker.hohmann@uni-oldenburg.de

# ABSTRACT

Both harmonic and binaural signal properties are relevant for auditory processing. To investigate how these cues combine in the auditory system, detection thresholds for an 800-Hz tone masked by a diotic (i.e., identical between the ears) harmonic complex tone were measured in six normal-hearing subjects. The target tone was presented either diotically or with an interaural phase difference (IPD) of 180° and in either harmonic or "mistuned" relationship to the diotic masker. Three different maskers were used, a resolved and an unresolved complex tone (fundamental frequency: 160 and 40 Hz) with four components below and above the target frequency and a broadband unresolved complex tone with 12 additional components. The target IPD provided release from masking in most masker conditions, whereas mistuning led to a significant release from masking only in the diotic conditions with the resolved and the narrowband unresolved maskers. A significant effect of mistuning was neither found in the diotic condition with the wideband unresolved masker nor in any of the dichotic conditions. An auditory model with a single analysis frequency band and different binaural processing schemes was employed to predict the data of the unresolved masker conditions. Sensitivity to modulation cues was achieved by including an auditory-motivated modulation filter in the processing pathway. The predictions of the diotic data were in line with the experimental results and literature data in the narrowband condition, but not in the broadband condition, suggesting that across-frequency processing is involved in processing modulation information. The experimental and model results in the dichotic conditions show that the binaural processor cannot exploit modulation information in binaurally unmasked conditions.

# I. INTRODUCTION

Auditory scene analysis (ASA) allows humans to detect, identify and track sound sources (e.g., talkers) in complex acoustic environments (e.g., Bregman, 1994). According to Bregman (1994), ASA partly relies on the grouping of auditory information that likely stems from the same sound source into single auditory objects. Important signal properties for auditory grouping are binaural and harmonic cues (e.g., Hukin and Darwin, 1995; Darwin and Hukin, 1999). Binaural information (interaural time differences, ITD and interaural level differences, ILD) allows the azimuthal localization of a sound source and constitutes a strong auditory grouping cue. Darwin and Hukin (1999), e.g., found that small differences in ITDs alone can be used to separate words in the absence of talker or fundamental frequency (F0) differences. Harmonicity is also a strong grouping cue that fuses individual components of a complex tone or speech formants with a common fundamental frequency (F0) into a single auditory object (e.g., Moore *et al.*, 1985; Hukin and Darwin, 1995). The human auditory system is sensitive to "mistuning", i.e., deviations from the harmonic frequency relationship between complex components. Moore *et al.* (1986) and Hartmann *et al.* (1990) have shown that deviations between 0.5-4% of the fundamental frequency can be detected and lead to the perception of a mistuned component as second auditory object in addition to the complex tone.

For a speech signal embedded in realistic acoustic environments, harmonicity and binaural cues of the talker co-vary with its spatial position and with those of interfering signals. The computational ASA models of Kepesi *et al.* (2007) and Ma *et al.* (2007) therefore explicitly assume combined processing of the two cues. The mechanisms of combining harmonic and binaural cues employed in these model studies have been shown to be effective in performing ASA and have been optimized using signal-processing techniques. It is unclear, however, to what extent they simulate auditory processing in a realistic way.

This study aims at further investigating possible auditory combination mechanisms by contributing psychophysical data and auditory model simulations on the detection of single sinusoidal components masked by a harmonic complex tone masker. This condition is simplified compared to realistic speech-in-noise scenarios in that the detection task is restricted to a single frequency component. The measurement paradigm therefore mainly allows conclusions about the processing mechanisms in single frequency channels close to masked threshold, which is a prerequisite for subsequent studies of information integration across frequency and of supra-threshold processing in models of ASA.

Several studies address auditory processing of combined binaural and harmonicity information: McDonald and Alain (2005) measured event-related brain potentials (ERPs) for harmonic and mistuned target tones in a ten-tone complex, while presenting the target tone either on the same or on a different loudspeaker than the masker (i.e., the other nine harmonics). Their behavioral and electrophysiological data showed that both harmonicity and location are evaluated to separate sounds and that localization cues can be used to resolve ambiguity in harmonicity cues. They found some evidence that harmonicity-based segregation of sound sources occurred during active and passive listening, whereas location effects were only observed during active listening. Based on this finding, they conclude that the evaluation of localization information is more reliant on active top-down processing than harmonicity processing. Their setup, however, did not allow conclusions as to which underlying binaural cues and processing schemes are responsible for their findings, as the stimuli at ear-level are difficult to control in a free-field experiment. Klinge *et al.* (2011) also investigated the combined influence of binaural and harmonic signal cues in the free sound field by measuring detection thresholds of a sinusoidal target component in a harmonic complex-tone masker. The target component was either in a harmonic or in a mistuned relationship to the masker with fundamental frequency F0 and could additionally be presented on a separate loudspeaker located at 90° azimuth. They found an addition of the two masking release effects: Both mistuning and spatial separation of the target decreased its detection threshold. As in McDonald and Alain (2005), however, the free-field setup employed by Klinge *et al.* (2011) does not exclude the exploitation of spectral cues (e.g., Hebrank and Wright, 1974) or ILD-based localization information.

Further evidence regarding the joint processing of binaural and harmonic cues was provided by Krumbholz *et al.* (2009), who used headphone experiments that allowed for a strict control of interaural parameters and could thus yield more precise findings on ITD processing. They found that subjects had difficulties to perform musical interval recognition (MIR) tasks with binaurally unmasked complex tones, i.e., the target intervals were only audible when listening with both ears, due to binaural release from masking. With increasing fundamental frequency, they found decreasing MIR performance. Since their complex tones were unresolved, the authors suggested that the temporal envelope fluctuations that would convey the required periodicity information were not accessible to the auditory system in binaurally unmasked conditions. According to Krumbholz *et al.* (2009), this can be described by a processing scheme in which the binaural processor is followed by a temporal integration stage that simulates binaural sluggishness (e.g. Grantham and Wightman, 1979; Kollmeier and Gilkey, 1990) and corrupts temporal information available to the subsequent pitch processor, leading to the observed MIR performance decrease.

Further studies directly investigated the relation between amplitude modulation in frequency subbands and binaural processing. Epp and Verhey (2009a) studied the combination of comodulation masking release (CMR, e.g., Hall *et al.,* 1984; Hall III *et al.*, 1990) and binaural masking level differences (BMLD, e.g., Licklider, 1948; Jeffress et al., 1956) in headphone experiments. Here, only interaural phase disparities were available as binaural cues. Epp and Verhey (2009a) found a linear addition of the two masking releases, which indicates a serial, rather than a parallel processing of CMR- and BMLD-related cues. Hall III et al. (2011) examined monaural and binaural masking release in CMR conditions and found data to be compatible with serial mechanisms of binaural and monaural masking release. However, data indicated that the combined masking release (BMLD and CMR) is smaller than the sum of BMLD and CMR. Nitschmann and Verhey (2012) measured BMLDs as a function of frequency separation between masker and the pure tone target signal. They found that BMLDs decrease with increasing spectral distance between masker and target and state that the observed decrease in masking release could be caused by modulation information not being available to the binaural system, but to the monaural system only. Thompson and Dau (2008) found that modulation frequency tuning is wider in the binaural than in the monaural domain, requiring broader modulation filters than monaural modulation filters as proposed in Dau *et al.* (1996). They speculate that such filters could be employed either before or after binaural processing.

In summary, the literature is somewhat inconclusive regarding the combination mechanisms of modulation and binaural processing. Three different general types of processing seem possible:

- Binaural processing precedes modulation processing
- Modulation processing precedes binaural processing
- Parallel processing stages, where the binaural stage has no or reduced modulation selectivity compared to the monaural stage

To further evaluate these hypotheses, this study investigated the combined influence of harmonic and binaural signal cues by psychophysically measuring detection thresholds of a single 800-Hz target tone in a resolved (fundamental frequency F0 = 160 Hz) or unresolved (F0 = 40 Hz) tone-complex masker. The target tone could either be harmonic or mistuned in relation to the masker, and was presented diotically ($M_0S_0$) or dichotically with an interaural phase difference of 180° ($M_0S_\pi$). For full control of the binaural stimulus parameters at ear-level and to exclude ILD and spectral cues, the measurements were performed with headphones. Psychophysical data were compared to predictions from a combination of established auditory processing models for the unresolved case, where harmonicity perception is driven by temporal modulation information that is known to be exploited for pitch analysis and mistuning detection (Moore *et al.*, 1985; Hartmann *et al.*, 1990; Lee and Green, 1994). In particular, the ability of modulation filters as proposed by Dau et al. (1996) to predict harmonicity-related release from masking in combination with a binaural processor that predicts binaural release from masking using a simple equalization-cancellation approach similar to Durlach (1963) was investigated for the three processing hypotheses.

## II. METHODS

### A. Subjects

Six normal-hearing listeners (4 male, 2 female) aged 24 to 32 years participated in the experiments. Their hearing threshold was below 15 dB HL in the range between 250 Hz and 4 kHz, as confirmed by standard audiometry. Before data acquisition, all subjects took part in one hour of training with the same stimuli as used in the final experiment. Five of the subjects received compensation on an hourly basis for their participation. The other subject was an author of this study (MK). The subjects provided written informed consent before participating in the experiments. The study protocol and consent procedure were approved by the Ethics Board of the University of Oldenburg and acknowledged by the German Research Council (DFG).

### B. Apparatus and stimuli

The experiments were conducted in a double-walled, sound-attenuating booth. The stimuli were generated digitally with a sampling frequency of $f_s$ = 48 kHz at runtime, on a PC using MATLAB. After conversion to analog signals by an external RME ADI-8 PRO D/A converter connected to a 24-bit RME DIGI96/8 PAD sound card, the stimuli were presented via Sennheiser HD 650 headphones. A Tucker Davis HB7 headphone buffer was used to drive the headphones. The subjects gave responses using a computer keyboard or mouse. Visual feedback was given on a computer monitor.

The stimulus configuration was identical to Klein-Hennig *et al.* (2012) and is illustrated in Figure 1. The target tone consisted of a pure-tone with a frequency $f_t$ of 800 Hz. The masker was a harmonic complex consisting of four harmonics below and four above the target tone. In Experiment 1, the masker had a fundamental frequency F0 of 160 Hz, such that the components had frequencies of 160, 320, 480, 640, 960, 1120, 1280 and 1440 Hz. In Experiment 2, the masker F0 was 40 Hz, leading to component frequencies of 640, 680, 720, 760, 840, 880, 920 and 960 Hz. The number of components was kept the same in the two experiments, which led to the fundamental being present in the first, but not in the second experiment. Both factors, the number of components and the presence or absence of the fundamental may have an influence on the processing of the target tone, and we had to decide which one to keep constant. To test the influence of the number of components, Experiment 3 used the same configuration as in Experiment 2, with 12 additional masker harmonics below and above the target component frequency, leading to a broad-band masker with 32 components ranging from 160 to 1440 Hz. The tones were added up in the temporal domain, with random phases identical on both channels (diotic, $M_0S_0$). To achieve binaural masking release, a dichotic $M_0S_\pi$ was created by applying an additional phase shift of 180° to the target tone on the right ear channel prior to the addition of the masker. Mistuning was applied by increasing the fundamental frequency F0 of the masker by a certain percentage, while keeping the target frequency $f_t$ constant. The mistuning values were 3.12 % in Experiment 1 and 2.64 % in Experiments 2 and 3. These values were chosen because they produced the largest masking releases in Klein-Hennig *et al.* (2012). The percentages lead to a difference of 20 Hz between the 4[th] masker component and the target component. In contrast to Klein-Hennig *et al.* (2012), all thresholds were measured while presenting a continuous, binaurally uncorrelated low-pass noise (380-Hz cut-off frequency, 4[th] order butterworth filter) at a level of 45 dB SPL to prevent the possible influence of cochlear distortion products by pure-tone harmonics (e.g., Goldstein, 1967; Pressnitzer and Patterson, 2001).

The stimuli had a duration of 400 ms, including on- and off-gating using 25 ms Hann windows.

## C. Procedure

The detection thresholds were determined using an adaptive 3-interval, 2-alternative forced-choice procedure. A 1-up, 2-down tracking rule was used, estimating the 70.7 %-correct point on the psychometric function (Levitt, 1971). The first interval always contained a reference stimulus and could not be selected as a response. The test subject had to indicate which interval contained the target tone. As in Klein-Hennig *et al.* (2012), the target signal was initially presented at a level of 65 dB SPL and was varied in 5 dB steps. The step size was reduced to 2 dB after the second and 1 dB after the fourth reversal. The masker level was fixed at 65 dB SPL. A new set of random phases was generated for each interval of each trial.

After eight reversals with 1dB steps, each adaptive run was terminated. The individual mean thresholds were calculated by arithmetically averaging over the last eight reversals of five experimental runs. The mean thresholds reported in the results are arithmetic averages over the individual averages of all subjects.

## D. Models

This section describes the single-channel model used for simulating the unresolved conditions from Experiments 2 and 3. Experiment 1 was excluded from modeling, as all masker components were outside the critical bandwidth of the on-target auditory filter. The development and evaluation of a multi-channel model employing across-frequency processing to account for this resolved condition is beyond the scope of this study.

## 1. Peripheral processing

To simulate auditory processing in the cochlea, a single-channel auditory preprocessing stage was implemented, similar to Dietz *et al.* (2009). It consisted of

- Real-valued 4th-order gammatone filtering at the target frequency $f_t$ = 800 Hz, with a bandwidth of one ERB,
- Half-wave rectification,
- 5th-order 770-Hz lowpass filtering (Breebaart *et al.*, 2001).

## 2. Modulation and binaural processing stages

The modulation processing stage used in the model is based on the envelope power spectrum model (EPSM) by Ewert and Dau (2000). It is used to extract modulation information that is used for target detection in the subsequent decision stage.

To extract the envelope of a stimulus, the time signal after peripheral processing was sampled down to a sampling frequency of 2000 Hz. Then, a single modulation filter (Dau *et al.*, 1997) was used for envelope processing. The modulation filter frequency was set to the fundamental frequency F0 = 40 Hz in harmonic conditions and to 20 Hz in the mistuned conditions. This value corresponds to the frequency difference between the fourth masker component and the target component. The frequency difference generates a 20-Hz beating that should be visible in the modulation spectrum and could be a detection cue. The 20-Hz modulation filter should extract this cue at the maximum SNR, leading to a favorable to signal-to-masker ratio that improves detection in the mistuned case, as opposed to the harmonic case. By performing simulations with a set of modulation-filter center frequencies in the range between 2 Hz and 50 Hz it was confirmed that the filter at 40 Hz (harmonic conditions) and 20 Hz (mistuned conditions) provided the largest detection cue, i.e., the simulation provides the lowest possible detection thresholds using these frequencies, and a multi-channel approach is not required here.

The model was divided into a monaural and a binaural pathway. For monaural processing, the absolute left and right ear internal signals after modulation processing were combined into a single signal by addition. The binaural processing stage is a simplification of the equalization-cancellation model as proposed by Durlach (1963) and calculates the absolute value of the difference between the left and right internal signals, which depend on the processing order (see section II.D.4 for the different versions of the binaural pathway).

**3. Decision stage**

Because experimental dichotic thresholds were at least 3 dB better than the diotic thresholds in the unresolved conditions (see section III), a combination of information across parallel monaural and binaural pathways, e.g., via a combination of sensitivity indices d', does not give a benefit compared to using the most sensitive pathway only. The target and reference intervals were therefore processed by the monaural pathway for the diotic and with the binaural pathway for the dichotic stimulus conditions, i.e., the respective most sensitive pathway was used. As a decision variable, the squared output signal of the employed pathway was integrated across the central 50% steady-state part of the interval, yielding a single value for each target and reference interval that establishes the decision variable and corresponds to its total energy $E_{total}$. Passing through the same adaptive procedure as the test subjects, the model chose the interval with the largest $E_{total}$. To limit detection accuracy, the decision variable was corrupted by adding a normally-distributed noise signal with zero mean to the output signal prior to integration in both pathways. The standard deviation of the noise distribution was set for the model to match the harmonic detection thresholds observed in the human data after 100 adaptive runs. This resulted in a fixed standard deviation of $\sigma_m = 0.01$ for the monaural pathway in all experiments. The standard deviation $\sigma_b$ for the binaural pathway was set up in the same way, to correctly predict the harmonic, dichotic threshold. $\sigma_b$ was also constant across experiments, but had to be adapted to each configuration of the binaural pathway, as the value range of the decision variable, i.e. the total energy of the signal, differed between the different processing orders. Respective values are given in section IV (model results). With this setup, $\sigma_m$ controlled the detection performance in the harmonic, diotic case, while $\sigma_b$ set the threshold for the harmonic, dichotic case. This way, the model simulations predict the data of the inharmonic conditions without any further free parameter, which can be directly compared to the human data.

### 4. Model configurations

Figure 2 illustrates the model configurations used to predict the data. The left panel shows the monaural processing pathway which is employed to predict the diotic data. This pathway was identical in the three model configurations that were tested. The model configurations differ in the order of binaural and modulation processors in the binaural processing pathway. In the first model configuration (second panel of Figure 2), binaural processing precedes modulation processing, hence modulation information is extracted from the output of the binaural processor. The second configuration (third panel of Figure 2) reverses this order: a common modulation stage processes left and right ear signals, followed by a binaural processor that processes the output of the modulation stage. The third model configuration (fourth panel of Figure 2) does not include modulation processing, simulating a binaural pathway without access to modulation information. Each simulated threshold was the average of five simulation runs with random phase angles per trial, as in the listening tests. To provide an estimate of model threshold variability, which is mainly due to the external noise induced by the random phases as well as the internal noise imposed on the detection variable, 20 simulated thresholds were computed for each condition.

## III. EXPERIMENTAL RESULTS

### A. Experiment 1: F0 = 160 Hz, $f_t$ = 800 Hz

The results of Experiment 1 are shown in the left panel of Figure 3. Experiment 1 used a fundamental frequency F0 of 160 Hz and a target frequency $f_t$ of 800 Hz ($5^{th}$ harmonic of F0=160 Hz). The number of masker components was $n_c = 8$. In the diotic $M_0S_0$ condition, a threshold decrease by 6.3 dB can be observed comparing the harmonic and the 3.12% mistuning condition. Presenting the target dichotically with an IPD of 180° lowered the thresholds from -12 to -17 dB in the harmonic (0% mistuning) condition, which corresponds to a BMLD of 5.5 dB. With a mistuning of 3.12%, the BMLD decreased from 5.5 to 0.7 dB. A 2-way repeated measures ANOVA showed a significant main effect of mistuning (p=0.004, F(1,5)=26.9) and IPD (p<0.001, F(1,5)=87.4) and a significant interaction effect between both factors (p=0.012, F(1,5)=15.0). Post-hoc Bonferroni-corrected pairwise comparisons showed a significant effect of mistuning in the diotic case (p=0.001), but not in the dichotic case (p>0.05). The effect of IPD was significant for the harmonic case (p<0.001), but not for the mistuned case (P>0.05).

### B. Experiment 2: F0 = 40 Hz, $f_t$ = 800 Hz

In this experiment, F0 was set to 40 Hz while keeping the target frequency $f_t$ at 800 Hz ($20^{th}$ harmonic of F0=40 Hz), to generate unresolved harmonics in the auditory filter centered at $f_t$. As in Experiment 1, the number of masker components was $n_c = 8$. The results are shown in the center panel of Figure 3. In the diotic condition, mistuning led to a decrease in detection threshold by 5.8 dB. In the harmonic 0% condition, a BMLD of 8 dB was found. At 2.64% mistuning, the BMLD decreased to 2 dB. A 2-way repeated measures ANOVA showed a significant main effect of mistuning (p=0.017, F(1,5)=12.4) and IPD (p<0.001, F(1,5)=54.2) and a significant interaction effect between both factors (p=0.005, F(1,5)=23.8). Post-hoc

Bonferroni-corrected pairwise comparisons showed a significant effect of mistuning in the diotic case (p=0.009), but not in the dichotic case (p>0.05). The effect of IPD was significant for the harmonic case (p<0.001) and for the mistuned case (P=0.025).

### C. Experiment 3: F0 = 40 Hz, $f_t$ = 800 Hz, broadband

This experiment used the same frequency configuration as Experiment 2 (F0 = 40 Hz, ft=800 Hz), but increased the number of masker components to $n_c$ = 32. This way, the masker had the same frequency span as in Experiment 1.

The results are shown in the right panel of Figure 3. In the diotic conditions, mistuning decreased the thresholds by 2.1 dB. In the dichotic conditions, a threshold decrease of 0.5 dB in the mistuned condition can be observed. Thus, the BMLD was 7 dB in the harmonic condition and 5 dB in the mistuned condition. A 2-way repeated measures ANOVA showed a significant main effect of IPD (p<0.001, F(1,5)=155,9), but no significant main effect of mistuning and no significant interaction effect between both factors (p>0.05). Post-hoc Bonferroni-corrected pairwise comparison showed a significant effect of IPD in the harmonic and the mistuned case (p<0.001).

## IV. MODEL RESULTS

Figure 4 shows the masking release by mistuning, i.e., the difference in masked thresholds between mistuned and harmonic condition (right and left data points of each panel in Fig. 3). The top row of panels shows data for Experiment 2 and the bottom row for Experiment 3 for the three tested model configurations. Gray symbols show individual human data, black symbols show mean and standard deviation across 20 simulated thresholds. Note that the same model was used in all diotic conditions. Predictions slightly vary because model runs were repeated for each condition. As described above, detection noise $\sigma_m$ was set so that the detection thresholds in the diotic harmonic condition matched the average experimental data.

The binaural noise $\sigma_b$ was set for each binaural model configuration so that the detection thresholds in the dichotic harmonic condition matched the average experimental data. This resulted in an average simulated BMLD in the harmonic condition of about 10 dB, which is only slightly higher than the experimental value of 8 dB. The model then predicted the masking release by mistuning in both the diotic and dichotic conditions, plotted in Fig. 4, without any further free parameter.

**A. Binaural processing before modulation processing**

In the processing chain of this model, the binaural equalization-cancellation stage processes left and right ear signals. The output is then analyzed by a modulation filter, as described in Section II.D.4 (see 2$^{nd}$ panel of Figure 2). The standard deviation of the internal noise was $\sigma_b = 0.04$. The predictions of this model configuration are shown in the left column of Figure 4. For Experiment 2 (upper left panel), the model shows a small 1 dB masking release by mistuning in the dichotic condition (right data point) that is not observed on average in the human data. Prediction variability, however, covers the range of data observed in the subjects. The diotic predictions (left data point) are in line with the human data, prediction variability covers the range of experimental data. Two of the subjects, however, show a relatively small masking release, indicating that they may be less effective in exploiting the mistuning cue than the model. In Experiment 3 (lower panel), the model predictions for the diotic and dichotic conditions are largely unchanged compared to Experiment 2, while in the subjects, masking release is reduced to smaller values between 0 dB and 4 dB in the diotic condition. The range of simulated data does not cover the range of experimental data in this condition.

**B. Binaural processing after modulation processing**

This model configuration performs modulation processing after auditory preprocessing. The binaural difference is calculated on the output of the modulation stage (see 3$^{rd}$ panel of Figure 2). The standard deviation of the internal noise was $\sigma_b = 0.0076$. The predictions are shown in the center column of Figure 4. For the dichotic condition of Experiment 2 (upper panel), the model predicts on average no masking release by mistuning, prediction variability covers the range of experimental data. The diotic predictions are in line with the human data. For Experiment 3 (lower panel), the model predicts more masking release for the diotic condition than observed in the subject data.

### C. Parallel processing (no modulation selectivity in the binaural pathway)

This model does not include a modulation filter in the binaural pathway (see 4th panel of Figure 2). The binaural processor directly processes the preprocessed signals. The model thresholds are shown in the right column of Figure 4. The standard deviation of the internal noise was $\sigma_b = 0.0161$. The model predicts no masking release by mistuning for Experiment 2 in the dichotic condition (upper panel), prediction variability covers the range of experimental data. The diotic predictions are in line with the human data. For Experiment 3 (lower panel), masking release by mistuning is again higher than in the subjects in the diotic condition. The dichotic model predictions are in line with the human data.

## V. DISCUSSION

### A. Psychophysical results

In summary, the results show that both harmonic and binaural manipulation of the stimuli affected the detection thresholds. All major effects were confirmed by a 2-way repeated measures ANOVA with subsequent Bonferroni-corrected post-hoc tests[1].

The observed diotic thresholds in the resolved case (Experiment 1) are in line with Klinge *et al.* (2011) and Klein-Hennig *et al.* (2012), who found a mistuning effect of the same magnitude using similar methods and similar frequency configurations (identical in the case of Klein-Hennig *et al.*, 2012). The diotic thresholds found for unresolved stimuli in Experiments 2 and 3 are in line with Klein-Hennig *et al.* (2012). Given that the thresholds in Klein-Hennig *et al.* (2012) were obtained without using a low-frequency masking noise as described in section II.B, cochlear distortion products caused by the pure-tone components appear to have little influence on the detection thresholds. This is in line with Oxenham et al. (2009), who found no effect of a low-pass masking noise, intended to mask distortion products, on the F0 difference limens for unresolved complexes.

---

[1] Significances were not changed when repeating the ANOVA without the author subject.

In all harmonic conditions, a significant BMLD of 5 to 7 dB was found. The resolved stimuli of Experiment 1 yielded a BMLD of 5 dB. The stimuli are comparable (with regard to resolvability and frequency range) to the 1000-Hz target-frequency conditions in Klinge *et al.* (2011), which yielded a BMLD of 5 dB as well. The unresolved conditions of Experiments 2 and 3 generated significant BMLDs of 7 dB, 2 dB more than the resolved condition in Experiment 1. With regard to resolvability, the results can be compared to the 8-kHz harmonic condition of Klinge *et al.* (2011), as a target frequency of 8 kHz and a fundamental frequency of 200 Hz leads to unresolved harmonics as well. Klinge *et al.* (2011), however, found a BMLD of 15 dB in that configuration, which is not in line with the results from Experiments 2 and 3. Since the stimuli in Klinge *et al.* (2011) were presented via loudspeakers, and the exploitation of binaural timing disparities at 8 kHz is highly limited, interaural level differences and spectral cues could have led to this large BMLD.

No significant effect of mistuning could be observed in the dichotic conditions of all masker configurations, which is a major finding of the present study. We trained the subjects well (one hour of training) and repeated the measurements at least five times per condition. Still, the subjects did not develop a sensitivity to mistuning in any of the dichotic conditions, which could mean that the binaural processor has no access to temporal modulation cues that convey mistuning information. This would be in favor of a parallel processing scheme, where the binaural pathway has only limited access to the periodicity information required for a detection advantage by mistuning at least for signals close to detection threshold. The data, however, do not rule out the possibility that the binaural processor may process modulation information for supra-threshold signals. The lack of a dichotic mistuning effect is not in line with Klinge *et al.* (2011), who still found a significant effect of mistuning in "separated" stimulus conditions where the target tone was presented from a loudspeaker located at 90° azimuth and the masker from a loudspeaker in front of the test subjects. However, the free-field setup does not exclude other localization cues such as the ILD and spectral cues, which possibly improved detection in their experiments.

As in Klein-Hennig *et al.* (2012), broad-band stimuli generated by additional masker components decreased the effect of mistuning to such an extent that no significant effect of mistuning could be observed. As the additional components were added outside of the passband of the auditory filter centered around the target frequency $f_t$=800 Hz, this hints at the involvement of across-frequency processing, as the available information within the critical bandwidth of the on-target filter should be identical in the narrowband and broadband conditions. In the mistuned broadband condition, the same BMLD is found as in the harmonic condition. The fact that only the diotic mistuned threshold is increased when adding further masker components, whereas the harmonic diotic and the two dichotic threshold remain unchanged, is noteworthy. This suggests that object binding mechanisms may compromise the detection of the mistuning in general while not having any influence on the binaural processing at threshold.

The masker plus target tone at threshold in our experiment is almost identical to a stimulus described by Duifhuis (1970) to have an audible pitch. Experiments by Lin and Hartmann (1997) suggested that different peripheral channels need to be combined to generate the "Duifhuis pitch". This conclusion was confirmed by Andermann et al. (2014) using MEG data, and their simulations suggest that nonlinear coupled cochlear filtering using a transmission line filterbank may be sufficient to provide the required interaction across frequency bands. The current data, however, suggest that nonlinear coupled peripheral filtering may not be the only mechanism, because it would affect all conditions, whereas we found only the diotic mistuned condition to be affected by band-widening.

To summarize, the mechanism of across-frequency processing of the cues remains somewhat unclear, and the current data challenge current models. Further physiological experiments using this type of stimuli may shed further light on this and may allow the development of models that contain physiologically meaningful mechanisms of across-frequency processing.

## B. Model results

The two model parameters (noise power in the monaural and the binaural pathway) were fitted to the harmonic conditions and were kept fixed for the mistuned conditions. This means that the masking release by mistuning in both the diotic and dichotic resolved conditions of experiments 2 and 3, as plotted in Fig. 4, was simulated without any free parameter.

On average, human data show no significant masking release by mistuning in the dichotic conditions of Experiments 2 and 3. However, there appears to be a small release from masking by mistuning in some individuals, similar to the model with the binaural processing preceding modulation filtering (left column of panels in Fig. 4). It thus may be that modulation selectivity is present in the binaural channel, but this does not lead to a pronounced masking release by mistuning. In the model, this is due to the noise in the binaural channel that masks most of the detection advantage by mistuning. Therefore, modulation selectivity leads to a pronounced threshold effect in diotic conditions, but not in dichotic conditions. Whether or not modulation selectivity in the binaural channel may play a role in supra-threshold signals remains to be investigated. Our data shows, however, that humans cannot take advantage of modulation selectivity in binaurally unmasked conditions close to threshold, which may apply to speech-in-noise conditions close to speech reception threshold.

The masking release by mistuning in the diotic condition of Experiment 2 (upper row of panels in Fig. 4) is simulated well by all model versions. In the broadband Experiment 3, however, all models show a larger release from masking by mistuning in the diotic conditionthan the humans. Thus, model predictions are not in line with the psychophysical result of a vanishing effect of mistuning with increasing masker bandwidth. To achieve correct detections in the broadband experiment, additional auditory filters for across-frequency processing would be needed. Epp and Verhey (2009b) successfully employed across-frequency processing for a model predicting the combined effects of CMR and BMLD, which could be a promising approach for the combination of mistuning and BMLD.

The simulations were restricted to modeling the mistuning effect by amplitude modulation filtering. Other models of harmonicity processing such as the harmonic template (Roberts, 2005) or harmonic cancellation (de Cheveigné, 1999; Deroche and Culling, 2011), could also be successful in modeling the mistuning effect in the diotic conditions and may be a better basis for modeling frequency integration. However, all these models would likely predict a mistuning effect in the dichotic conditions when combined with binaural processing in a serial way. The main finding that the binaural auditory system does not seem to have access to harmonicity cues at threshold is therefore unrelated to the specific choice of the harmonicity model.

## VI. CONCLUSIONS

- The psychophysical data show that release from masking by mistuning and binaural disparity do not combine in an additive way both in the resolved (Experiment 1) and unresolved (Experiments 2 and 3) conditions. In particular, masking release by mistuning was present in all diotic conditions, but was not present in any of the

- unresolved dichotic conditions. It was small and not significant in the resolved dichotic condition. This indicates that mistuning can provide a detection advantage, but not in binaurally unmasked conditions.
- The psychophysical data accentuate the importance of across-frequency processing in configurations where the target tone is resolved or where the harmonic masker complex is broadband, i.e., when a large number of harmonic masker components lie outside of the passband of the on-target auditory filter.
- The model results for the unresolved narrowband diotic condition show that modulation processing is able to account for release from masking by mistuning in this condition.
- Regarding the combination of binaural and modulation information required for detection, the model results show that a binaural processor without access to modulation information predicts the data of the unresolved dichotic conditions. Adding modulation selectivity to the binaural channel changed estimated detection thresholds only slightly. It can therefore not be ruled out that modulation selectivity exists in the binaural channel, but if so, it is not effective close to masked threshold.
- The data suggests that humans cannot take advantage of modulation selectivity in binaurally unmasked conditions close to threshold, which may apply to speech-in-noise conditions close to speech reception threshold.
- The modeling results form a basis for a subsequent investigation and modeling of combinatory effects in resolved as well as wideband unresolved harmonic complexes that, according to the data presented here, require across-frequency processing.

## ACKNOWLEDGMENTS

This study was supported by the DFG (SFB/TRR31 'The Active Auditory System') and by the European Union under the Advancing Binaural Cochlear Implant Technology (ABCIT) grant agreement (No. 304912). We thank the Medical Physics group and Birger Kollmeier for constant support and Astrid Klinge-Strahl, Stephan Ewert, and Georg Klump for fruitful discussions. We also thank two anonymous reviewers for their valuable comments.

# FIGURES

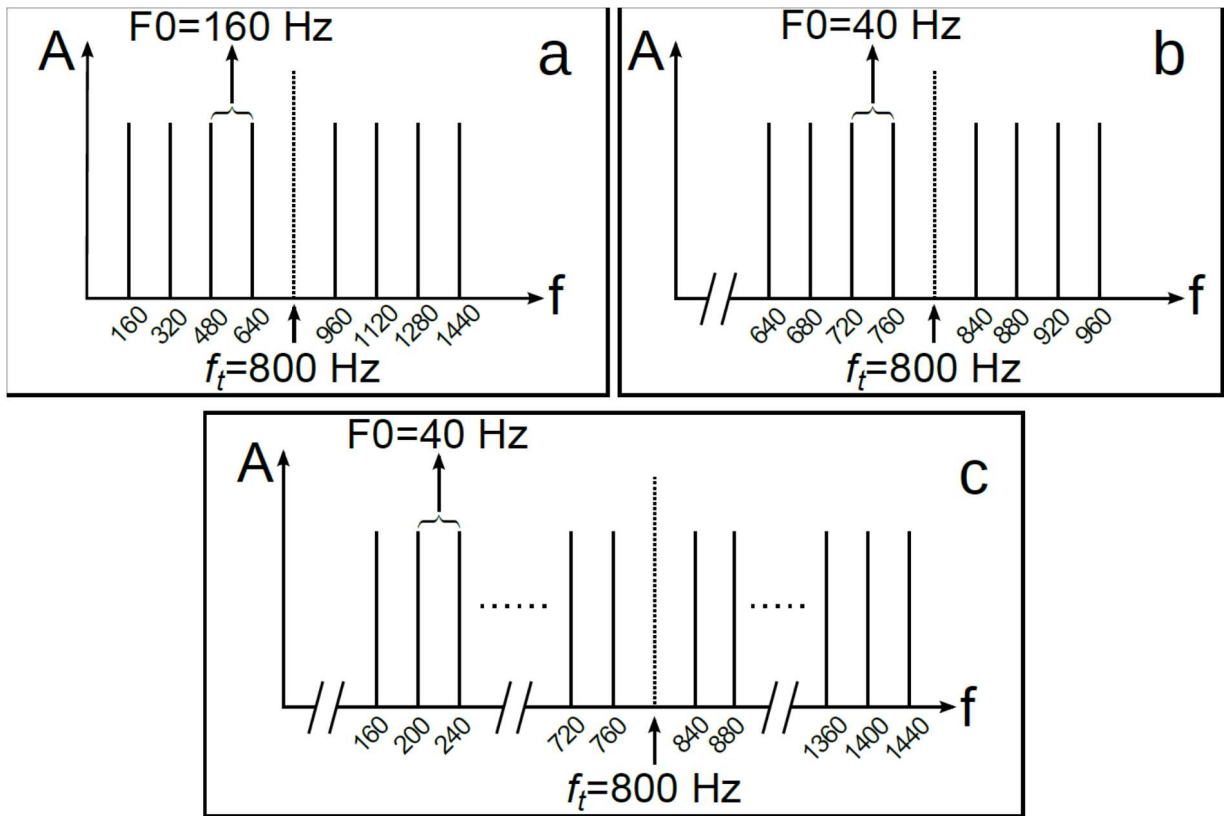

Figure 1: Pictograms of the frequency configurations for the stimuli used in the experiments. The masker harmonic complex is indicated by black lines, its respective fundamental frequency F0 is indicated in each panel. The dotted line represents the frequency $f_t$ of the target tone that had to be detected. Panels a and b indicate the configurations for Experiments 1 and 2. Panel c shows the configuration of the broadband stimuli with additional masker harmonics used in Experiment 3.

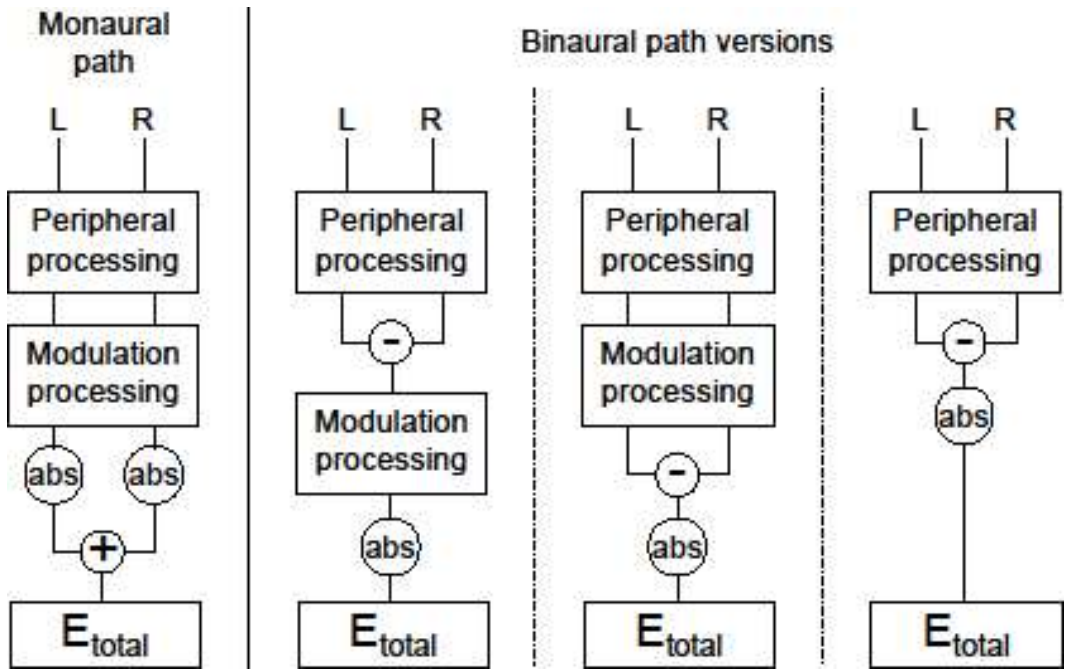

Figure 2: Schemes of the monaural pathway and the three tested binaural pathway configurations, employing different processing order. First panel: Monaural pathway employing modulation processing to predict diotic thresholds. Second panel: Binaural processing, indicated by the "-" precedes modulation processing. Third panel: Modulation processing precedes binaural processing. Fourth panel: Binaural processing has no access to the output of the modulation processor. The peripheral processing as well as the calculation of the integrated power $E_{total}$ of the respective output signal is identical in all configurations.

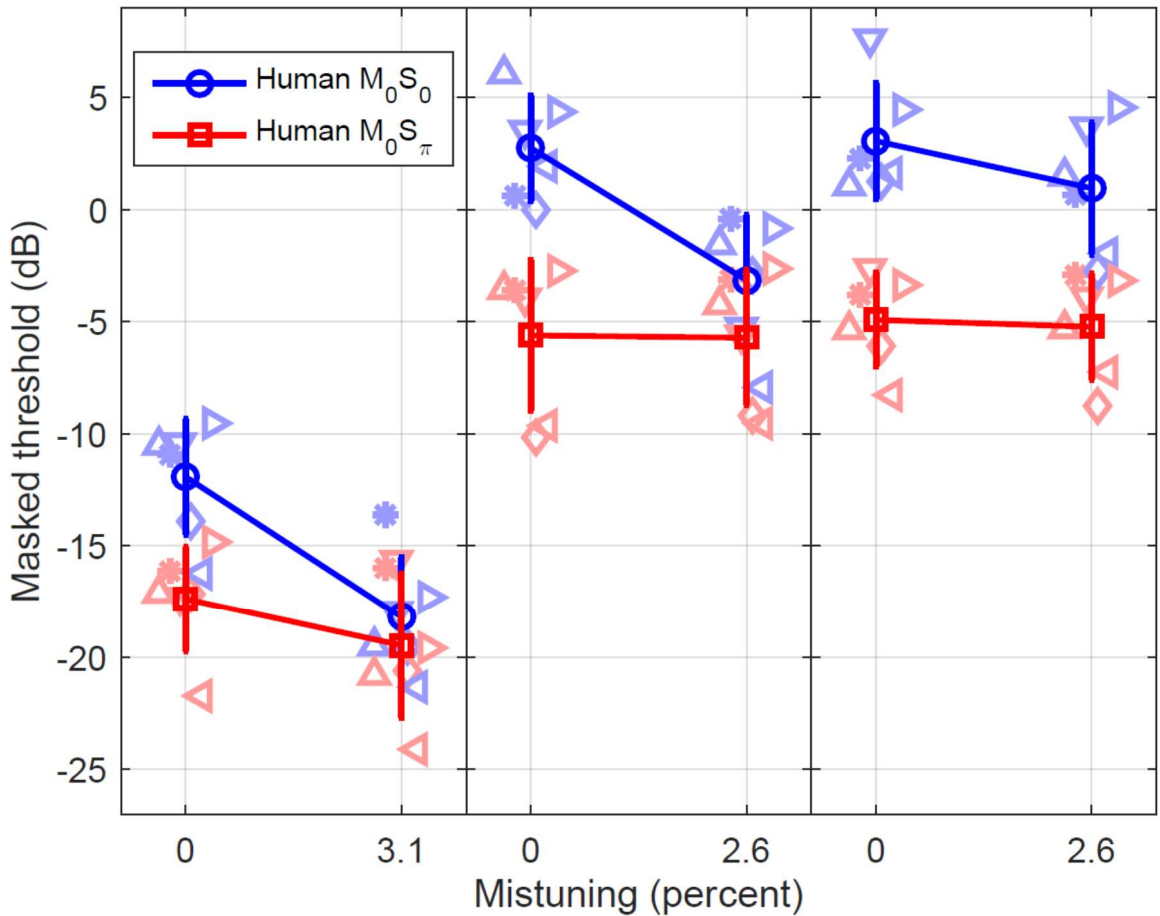

Figure 3: (Color online) Detection thresholds for Experiments 1 to 3 (left to right). Data points and error bars show mean and standard deviation across six subjects. Results for a diotic target tone ($M_0S_0$) are plotted with circles. Squares indicate thresholds for a target tone with an interaural phase difference of 180° ($M_0S_\pi$). The different symbols show individual data of the six subjects.

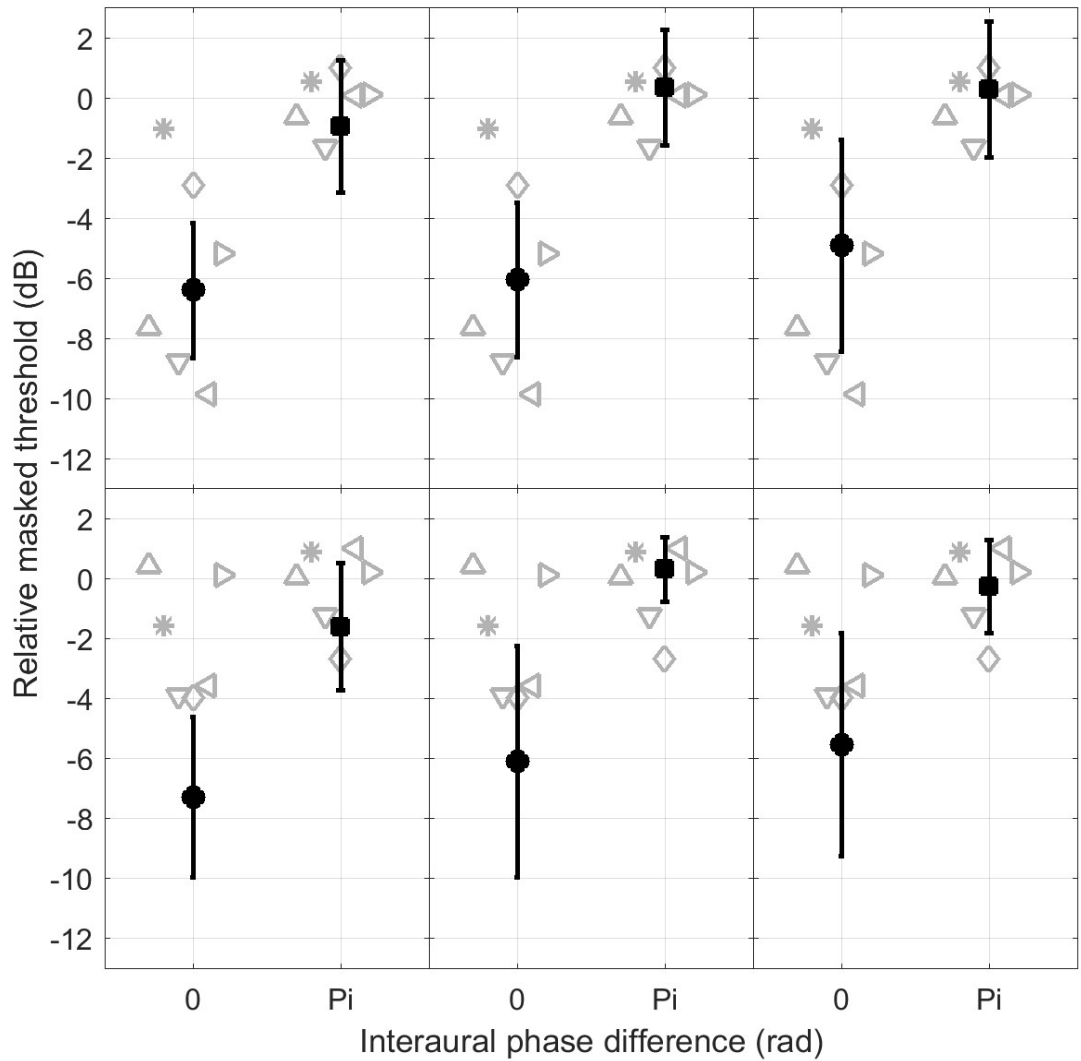

Figure 4: Masking release by mistuning (difference in masked thresholds between mistuned and harmonic condition) for Experiment 2 (top row of panels) and Experiment 3 (bottom row of panels) for the three tested model configurations (left column: binaural processing precedes modulation processing; center: modulation processing precedes binaural processing; right: binaural processing has no access to the output of the modulation processor). Gray symbols show individual human data, black symbols show model data (mean and standard deviation across 20 simulated thresholds).